# EPR measurements of ceramic cores used in the aircraft industry

Ireneusz Stefaniuk,
Iwona Rogalska,
Piotr Potera,
Dagmara Wróbel

**Abstract.** In this work the electron paramagnetic resonance (EPR) spectra of $Al_2O_3$ and $ZrO_2$ powders were measured for different size of grains (0.074, 0.044 mm) as well as the mullites (0.07 and 0.12 mm). Also were investigated the ceramic cores. The measurements were performed at room temperature and in the temperature range from 140 up to 380 K. The main purpose of this work was to investigate the possible relationships between the EPR spectra and the size of powder grains as well as the identification of EPR spectra in view of the potential application of EPR technique as a fingerprinting method.

I. Stefaniuk, I. Rogalska✉, P. Potera, D. Wróbel
Institute of Physics,
University of Rzeszów,
16a Rejtana Str., 35-310 Rzeszów, Poland,
Tel.: +48 17 872 1094, Fax: +48 17 872 1283,
E-mail: i.rogalska@if.univ.rzeszow.pl

## Introduction

The ceramic nanopowders are widely used in various industries, including the aerospace industry. The polymer nanocomposites stiffened by ceramic nanofilling are characterized by a very high hardness and resistance for abrasion in comparison with composites in a micrometric scale. The annealing of $Al_2O_3$ powders at temperatures 350, 600, 900°C do not influence the size of particles of the investigated powders. Heating at a temperature of 1200°C leads to a 30% growth of the average size of grains (due to the fritting processes) and, at the same time, the growth of crystallinity degree of alumina and the phase transition $\delta, \gamma, \eta, \varepsilon\text{-}Al_2O_3 \rightarrow \alpha\text{-}Al_2O_3$ takes place. The conditions of synthesis of the precursor as well as the usage of its modifier significantly influence the $Al_2O_3$ morphology. Also important is the reaction of environment in which the homogenization of both $Al_2O_3$ and nanometric $ZrO_2$ powders takes place. The environment, in which particles of both powders have the same electric charge signs, leads to the forming of mechanically resistant agglomerates. This is detrimental for the condensation of material during fritting. In the sintering process, the crack sizes about of hundreds of micrometers as well as inhomogeneity of packing in an individual micrometer scale are created [7].

Mullite is certainly one of the most prominent ceramic materials. It has become a strong candidate material for advanced structural and functional ceramics in recent years. The reason for this development is the outstanding properties of mullite – low thermal expansion, low thermal conductivity and excellent creep resistance. Mullite in the strict sense, with the composition $3Al_2O_3 \cdot 2SiO_2$, has been the subject of numerous crystal structure and crystal-chemical studies, not only for basic reasons but also because of its outstanding electrical, mechanical and thermal properties, and its wide use in traditional and advanced ceramics. The fundamental building unit in mullite is the chain of edge-sharing $AlO_6$-octahedra. A main advantage of oxide fiber/mullite matrix materials besides oxidation stability is the low-cost fabrication, especially in the case of porous mullite matrix composites. Because of the short duration of the thermal load during entry to the atmosphere (for re-entry space vehicles the total exposure time of the heat shield at a maximum temperature can be taken as less than 20 h) oxide fiber/mullite matrix materials are considered to be usable up to peak temperatures of 1500°C [5].

The aim of this work is to investigate by electron paramagnetic resonance (EPR) methods the role of cores and shapes of basic $Al_2O_3$ materials used for industrial applications. The motivation for this study comes from the need to solve the problem of fractures of ceramic cores and shapes.

We intentionally repeat investigation of oxide alumina $Al_2O_3$ and mullite $Al_6Si_2O_{13}$, but these are materials from different manufacturers than previously studied in early work [7]. The study was expanded to include new samples of silicon dioxide $SiO_2$, zirconium dioxide $ZrO_2$ and zirconium silicate $ZrSiO_4$, which are also used to obtain cores and ceramic molds. The main aim is to find any differences in the introduced doping, which may affect to the cracking form. During the research within the Centre for Advanced Technology AERO-NET – Aviation Valley, in the project 'Modern material technologies used in the aerospace industry' (http://pkaero.prz.edu.pl), has been detected that a large effect on the strength and durability of ceramic forms is the origin of the material. In spite of the measurements of X-ray powder from different manufacturers are not detected differences in the composition which could have an impact on cracking form. It was, therefore, decided to use the method of electron paramagnetic resonance (EPR), whose preliminary results allow us to detect differences in the investigated materials.

The results in this paper focus on the exact characterization of paramagnetic centres and the analysis of the relaxation times of these centres, which in the future may give an answer to the question why some form of break, while others do not, despite the same composition.

### Experimental details

For the experiment, the samples of corundum $Al_2O_3$, mullite and $ZrO_2$ with different size and incorporating a second phase were used. The ceramic core prepared materials by the high-pressure injection method, we also studied (Table 1). In previous work [7] it has been a mistake in determining the grain size, corected values of grain size of mullite should be 0.07 mm and 0.12 mm consequently.

For the EPR measurements, a standard X-band (~ 9 GHz) spectrometer, produced in Wrocław Technical University, with digital registration of the spectra was used. The temperature measurements were done using the digital temperature control system (Bruker

**Table 1.** The specification of the corundum samples

| Sample | $Al_2O_3$ 0.074 mm | $Al_2O_3$ 0.044 mm | $ZrO_2$ 0.044 mm | Ceramic core 1 | Mullite 0.07 mm | Mullite 0.12 mm |
|---|---|---|---|---|---|---|
| Qualitative analysis of phase composition by X-ray diffraction methods | α-$Al_2O_3$ – corundum 93.9 ± 0.1(%) $NaAl_{11}O_{17}$ – β-$Al_2O_3$ 6.1 ± 0.1(%) | α-$Al_2O_3$ – corundum 95.5 ± 0.1(%) $NaAl_{11}O_{17}$ – β-$Al_2O_3$ 4.5 ± 0.1(%) | ~ 100% | $SiO_2$ (120 mesh) – 37.5 mas.% $SiO_2$ (milled) – 37.5 mas.% $ZrSiO_4$ (300 mesh) – 13 mas.% $Al_2O_3$ (325 mesh) – 12 mas.% SILIPLAST HO – 25 mas.% | $Al_2O_3$ – 76.86% $SiO_2$ – 22.8% | $Al_2O_3$ – 75.04% $SiO_2$ – 24.5% |

ER 4131VT), which allows the temperature range from 100 to 500 K.

**Result and discussion**

The obtained EPR spectra are presented in Figs. 1 to 4. In spite of the similar chemical compositions of materials, differences between of EPR spectra were detected. The analysis of the line positions suggest that the lines with $g_{eff}$ = 4.28, $g_{eff}$ = 2.00 may be attributed to $Fe^{3+}$ ($S$ = 5/2) ions, because they present a typical spectrum for the so-called disordered systems [6, 8] present in the glassy hosts [9]. The line intensities decrease progressively showing the evolution of the relative line shapes and the intensities at $g_{eff}$ = 4.3 from isolated ions in local tetrahedral (and eventually octahedral) sites [9]. The line with $g_{eff}$ = 1.98 may be attributed to $Cr^{3+}$ ($S$ = 3/2) ions in the slightly distorted octahedral sites [2]. In this paper we would like to present results for the ceramic of dopants.

For samples no. 1 and 2, high differences in EPR spectrum was observed, as due to appearing of new line near 270 mT. This line is probably a result of the presence of $Cr^{3+}$ ($S$ = 3/2) ions in different size $Cr_2O_3$ nanoparticles [7].

In Fig. 1 marked lines are seen using arrows to the identification of paramagnetic centres. The wide line near 300 mT consists of a minimum of two components. They can be used for the analysis of changes in ion concentration and indirectly in the component materials. For the $ZrO_2$ (0.149 mm) sample the calculated $g_{eff}$ – factor values for each line are the following: $g_{eff}$ =

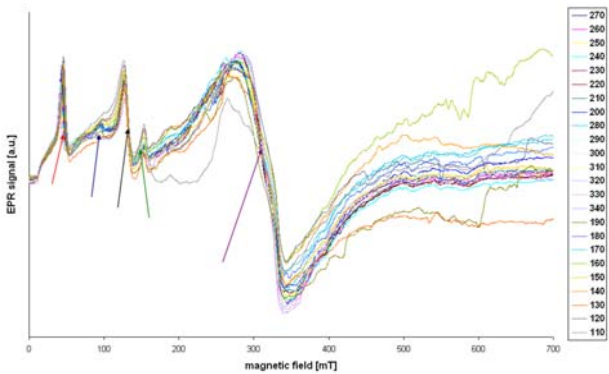

**Fig. 1.** The temperature dependence for ceramic core obtained from materials no. 1 (small sample).

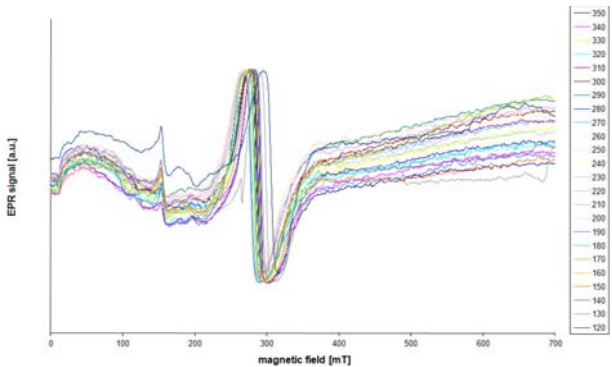

**Fig. 2.** The EPR spectra of $ZrO_2$ 0.044 mm at different temperatures.

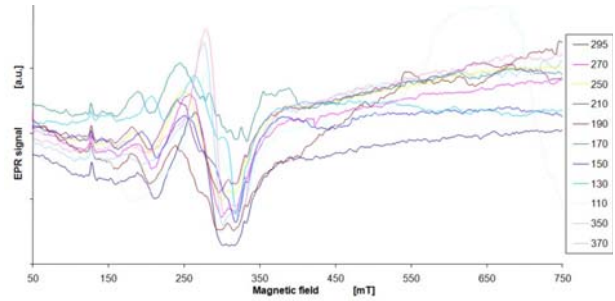

**Fig. 3.** The EPR spectra for $Al_2O_3$ powder (0.074 mm) at different temperatures.

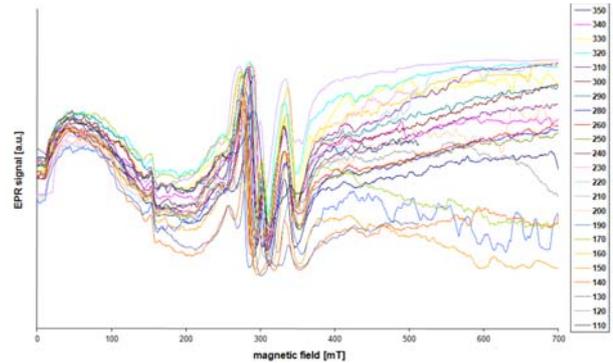

**Fig. 4.** The temperatures dependence of EPR spectrum for mullite 0.07.

6.84; $g_{eff}$ = 4.23; $g_{eff}$ = 2.24. However, for the sample 0.044 mm we obtained the values: $g_{eff}$ = 6.82; $g_{eff}$ = 2.54; $g_{eff}$ = 2.37; $g_{eff}$ = 1.88. The analysis of the line positions suggest that the lines with $g_{eff}$ = 4.23 may be attributed to $Fe^{3+}$ ($S$ = 5/2) ions, analogously then for the $Al_2O_3$ sample. The estimation of the spin-lattice relaxation (Figs. 5, 6) time $T_1$ can be made using the conventional method of line broadening, using the expression [1, 3, 4]:

(1) $$T_1^{-1} = 2.8 \times 10^{10} \pi g \Delta B$$

where $\Delta B$ (in $T$) is the spin-phonon part of the EPR linewidth (it is a change in the line width with temperature). In the temperature range 140–370 K the relaxation time $T_1$ is governed by the Orbach process [1, 3, 4]

(2) $$T_1^{-1} = A \left( \exp(\delta/k_B T) - 1 \right)^{-1}$$

where $T$ is the temperature, $k_B$ is Boltzman constant,

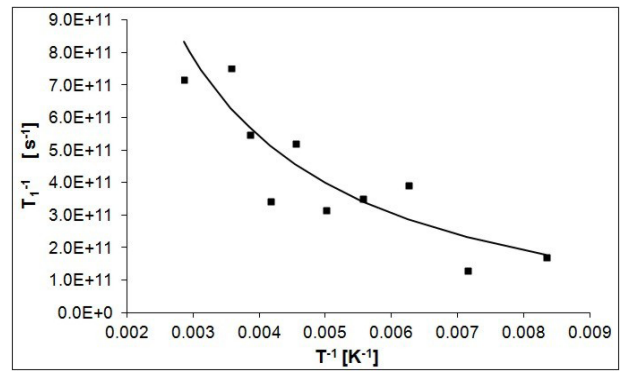

**Fig. 5.** Temperature dependence of the spin-lattice relaxation time $T_1$, of the mullite 0.12. The solid curve is an exponential fit to the obtained data by the equation (for $Fe^{3+}$ ions).

**Table 2.** The parameters of spectra for the mullite 0.12 ($B_{pp}$ – peak-to-peak line widths, $B_{res}$ – resonance field, $A$ – EPR signal amplitude, $g_{eff}$ – factors)

| Temp. (K) | Line 1 | | | | Line 2 | | | | Line 3 | | | |
|---|---|---|---|---|---|---|---|---|---|---|---|---|
| | $B_{pp}$ (mT) | $B_{res}$ (mT) | $A$ (a.u.) | $g_{eff}$ | $B_{pp}$ (mT) | $B_{res}$ (mT) | $A$ (a.u.) | $g_{eff}$ | $B_{pp}$ (mT) | $B_{res}$ (mT) | $A$ (a.u.) | $g_{eff}$ |
| 350 | 18 | 161 | 94.5 | 4.11 | 22 | 285 | 68 | 2.32 | 10 | 336 | 76.5 | 1.97 |
| 340 | 6 | 155 | 68.5 | 4.27 | 11 | 280.5 | 64 | 2.36 | 9 | 336.5 | 73.5 | 1.97 |
| 320 | 18 | 161 | 104.5 | 4.11 | 15 | 280.5 | 93 | 2.36 | 12 | 337 | 61.5 | 1.97 |
| 300 | 8 | 155 | 107.5 | 4.27 | 11 | 276.5 | 82 | 2.40 | 13 | 334.5 | 71 | 1.98 |
| 280 | 9 | 156.5 | 113 | 4.23 | 14 | 274 | 65.5 | 2.42 | 11 | 335.5 | 59.5 | 1.97 |
| 260 | 13 | 159.5 | 100 | 4.15 | 13 | 276.5 | 81.5 | 2.40 | 11 | 339.5 | 59 | 1.95 |
| 240 | 12 | 159 | 107 | 4.17 | 13 | 273.5 | 89.5 | 2.42 | 11 | 340.5 | 41 | 1.95 |
| 220 | 11 | 157.5 | 106 | 4.21 | 13 | 270.5 | 75.5 | 2.45 | 11 | 335.5 | 58.5 | 1.97 |
| 200 | 9 | 156.5 | 118.5 | 4.23 | 111 | 222.5 | 106 | 2.98 | 11 | 338.5 | 68 | 1.96 |
| 180 | 8 | 156 | 129.5 | 4.25 | 12 | 274 | 95 | 2.42 | 9 | 336.5 | 56 | 1.97 |
| 160 | 7 | 156.5 | 126 | 4.23 | 10 | 283 | 105.5 | 2.34 | 9 | 336.5 | 51 | 1.97 |
| 140 | 6 | 156 | 116 | 4.25 | 22 | 302 | 172.5 | 2.19 | 8 | 338 | 37 | 1.96 |
| 120 | 7 | 155.5 | 156.5 | 4.30 | 28 | 309 | 148.5 | 2.14 | 26 | 374 | 141.5 | 1.77 |

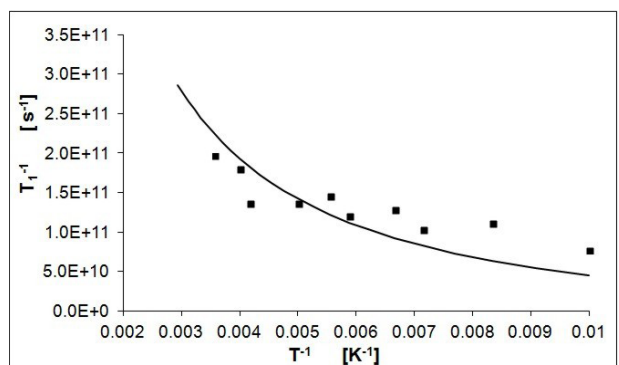

**Fig. 6.** Temperature dependence of the spin-lattice relaxation time $T_1$, of the ceramic core no. 1. The solid curve is an exponential fit to the obtained data by the equation.

parameter $\delta$ (cm$^{-1}$) represents the energy splitting between the ground paramagnetic centres state and the first excited state, whereas $A$ is a constant characteristic of the Orbach process (in s$^{-1}$) [1].

On the axis of the abscissa the inverse of temperature $T^{-1}$ (K$^{-1}$) is seen, on the axis of the ordinates the inverse of the spin-lattice relaxation time $T_1^{-1}$ (s$^{-1}$).

Orbach fit for ceramic core (resonant field line approximately 125 mT, for Fe$^{3+}$ and other ions), mullite (about 150 mT, for Fe$^{3+}$ ions).

## Conclusions

In order to improve the quality of ceramic molds and cores obtained from powders of $Al_2O_3$, $ZrO_2$ and mullite, it is necessary to keep a high degree of purity powders. Comparing the results for the same material shown in this and previous work [7], significant differences in the spectra of EPR have been observed. The EPR spectra for these samples differ quite considerably in shape, the intensity of individual components of the line, the width and the resolution and the signal-to-noise ratio, even though they are made using the same spectrometer EPR in the same conditions. These results correlate with the content of impurities in the powders from different manufacturers. Analyzing changes in line widths with the Orbach model for materials which are composed of different dopants, it can be detected of character line width changes even of the overlapping of EPR lines. By the EPR method we detected impurities whose standard X-ray diffraction (XRD) test are not noticable.

For $Al_2O_3$ powders from different parts of the some size of grain, different EPR spectra was obtained. As a result of the analysis, the identification of existing complexes of paramagnetic ions were performed, where nanoparticles $Cr_2O_3$ in addition to chromium and iron were detected. $Cr_2O_3$ phase occurs only in part II of $Al_2O_3$ powder with a particle size 0.074 mm, and a small amount in mullite powder.

The temperature dependence of the EPR line of the peak-to-peak ($B_{pp}$) linewidths were also measured. From these measurement, the values of the broadening ($\Delta B$) of the EPR line width, can be determined.

The analysis of the temperature dependence for the EPR line width in $Al_2O_3$, mullite and $ZrO_2$ powders and ceramic cores was performed with the Orbach process. Parameters obtained from the model of Orbach are presented in Table 3. For the mullite 0.12 mm and ceramic core adjustment is consistent with the Orbach process.

**Table 3.** Parameters obtained from the Orbach process

| | Ceramic core | Mullite 0.12 | Ref. [1] | Ref. [3] |
|---|---|---|---|---|
| $A$ (s$^{-1}$) | $16 \times 10^{10}$ | $45 \times 10^{10}$ | $37 \times 10^{10}$ | $10^{10}$ |
| $\delta$ (cm$^{-1}$) | 105 | 105 | 168 | 105 |

**Acknowledgment.** Financial support of Structural Funds in the Operational Programme – Innovative Economy (POIG) financed from the European Regional Development Fund – Project 'Modern material technologies in aerospace industry', is gratefully acknowledged. No. POIG.01.01.02-015/08-00.